# Influence of dispersion stretching of ultrashort UV laser pulse on the critical power of self-focusing

A.A. Ionin[1], D.V. Mokrousova[1,2*], D.A. Piterimov[1,2],
L.V. Seleznev[1], D.V. Sinitsyn[1], E.S. Sunchugasheva[1]

[1]P.N. Lebedev Physical Institute of the Russian Academy of Sciences, 53 Leninsky prosp., 119991 Moscow, Russia

[2]Moscow Institute of Physics and Technology, 9 Institutskiy Per., 141701 Dolgoprudny, Moscow Region, Russia

* daria.mokrousova@yandex.ru

The critical power of self-focusing in the air for ultrashort UV laser pulses stretched due to dispersion from 90 up to 730 fs was experimentally measured. It was shown that the pulse duration enhancement due to its propagation in condensed media within above mentioned time-range leads to almost linear decrease of self-focusing critical power. It was also observed that when the pulse peak power exceeds the critical one, the maximum of linear plasma distribution along the UV laser filament does not shifted in the direction opposite to the laser pulse propagation as in the case of IR laser filament, but remains at the geometrical focus.

## 1. Introduction

The energy of ultraviolet (UV) photon is considerably higher than that of infrared (IR) one. As a result, a probability of multiphoton ionization is much higher for UV laser pulse than for IR one. Therefore, ultrashort UV laser pulses are preferable for some applications, for instance, extended plasma channels formation during filamentation [1], laser triggering and guiding of long electric discharge [2, 3] or laser-induced breakdown spectroscopy [4]. Tripling the frequency of emission of Ti:Sa laser systems or doubling the frequency of emission of dye lasers and their amplifying in excimer laser amplifiers [1, 5-7] are the most generally implemented procedures of obtaining high-power UV pulses. In these optical schemes, laser pulse propagates through a large volume of pass-through optics (thick optical windows of gas laser amplifiers, spatial filters, etc.) and an active media. Because of the dispersion, pulse duration increases differently for various schemes in dependence on pass-through optics material and its thickness. The amplified pulse duration can be measured by means of an autocorrelator based on $BaF_2$ crystal [8] or gas cell filled with XeF [9]. Due to the pulse dispersion stretching, the contribution of inertial Kerr effect to nonlinear refractive index increases [10, 11]. This factor and phase modulation in pass-through optics [12] leads to enhancement of effective nonlinear refractive index and a decrease of critical power of self-focusing:

$$P_{cr} \sim \lambda^2 / n_0 n_2 \qquad (1).$$

It is vital to know the critical power of self-focusing in the air for experiments on nonlinear propagation of high-power laser pulses through the atmosphere in filamentation regime, including multiple filamentation. In this paper, we experimentally evaluate a dependence of the critical power of self-focusing on dispersion stretching of UV ultrashort pulse in the air.

## 2. Experiment and discussion

In the experiment we used pulses of third harmonics of Ti:Sa laser system at the central wavelength of 248 nm. Initial pulse duration after the third harmonics generator was 90 fs. Pulse propagation through high-power laser system optics was modeled by passing the beam through fused silica plane-parallel optical elements of different thickness. Pulse duration was controlled by an autocorrelator based on gas cell filled with XeF [9]. Pulse energy could reach 200 μJ and was varied with a diffraction attenuator. The laser beam was focused in the air using a spherical mirror with the focal length of 1 m. Linear plasma density distribution along the beam propagation was measured by a system, described in [12, 13]. Fig. 1 shows plasma density distribution obtained for different pulse peak powers for the pulse duration of 450 fs. Zero mark at the x-axis corresponds to the geometric focus position. Because of high UV quantum energy (5 eV), plasma formation in the focal waist occurred even at low subcritical pulse powers. Pulse power growth resulted in extending an area of the pulse propagation, where laser intensity was sufficient to effectively ionize the air. This can be observed as an increase of linear plasma density around the waist and symmetrical elongation of the plasma channel (fig.1, 52 – 80 MW). When the pulse power exceeded the critical one, plasma channel elongation became asymmetrical relatively to the geometrical focus. In this case, the beginning of the channel (nonlinear focus) shifts towards the pulse source considerably as compared to the channel extension behind the linear focus (fig.1, peak power is higher than 90 MW). As it was shown in [12], a laser beam propagated through transparent condensed medium is undergone by additional dynamical phase modulation, which can be considered as another non-stationary Kerr lens with the focus length $f_{nl\_solid}$, and the total pulse focusing can be calculated as

$$f^{-1}(\tau) = f_0^{-1} + f_{nl\_solid}^{-1}(\tau) + f_{nl\_air}^{-1}, \qquad (2)$$

where $f_0$ – is geometric focus and $f_{nl\_air}$ – Kerr self-focusing in the air. So, we have two nonlinear processes influencing upon the nonlinear Kerr index and plasma formation beginning: dynamic phase modulation in a silica optical element introduced into the beam path, and pulse Kerr self-focusing in the air. For given pulse duration fused silica elements were placed before the attenuator, therefore dynamic phase modulation was constant during energy variation. Thus, measured critical power at fixed pulse duration and phase modulation was associated only with Kerr self-focusing of the beam in the air (fig.1). Different pulse durations corresponded to different lengths of the fused silica elements and, consequently, self-modulation in condensed media. Therefore, the dependence of critical power of self-focusing in the air on pulse duration was obtained taking into account phase modulation in condensed medium. Considered processes take place in real high-power UV laser systems mentioned above.

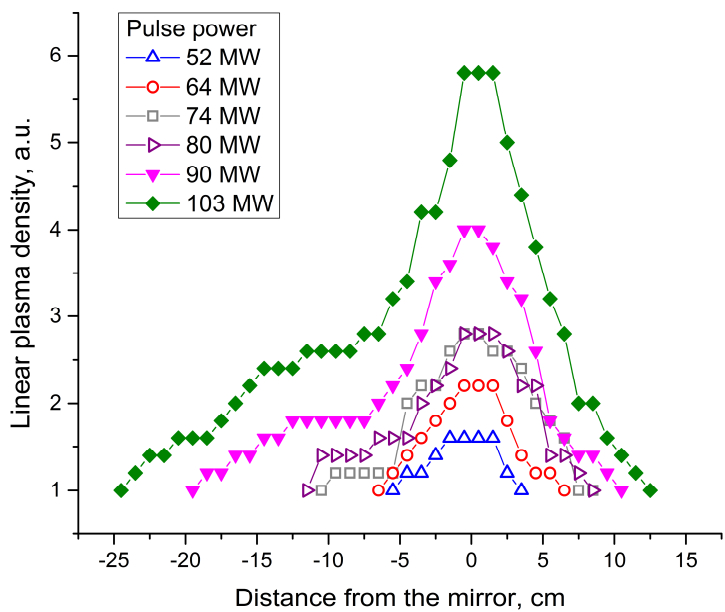

Fig.1. Linear plasma density distribution along the beam propagation obtained for different UV pulse powers. Pulse duration is 450 fs.

It is worth noticing that in contrast to IR filamentation [11], for UV pulses the position of the maximum for linear plasma density distribution remained in the vicinity of the geometrical focus. Apparently, the most intense slice of the pulse (both in time and beam cross-section) is collapsed in the nonlinear focus (a start of the channel) [12], whereas whole pulse periphery is focused in the geometrical focus [14]. Pulse periphery contains a substantial part of the pulse energy, so because of the high energy of UV photon a noticeable plasma channel occurs in the linear focus.

Linear plasma density distributions for pulses with duration from 90 to 730 fs were experimentally obtained using the same method. From these graphs, we evaluate the critical power of self-focusing as a power, corresponding to asymmetrical plasma channel elongation. The dependence of the critical power of self-focusing on the pulse duration is presented in fig.2. The dependence has predictable monotonically decreasing character close to linear one: the critical power decreases from 100 MW at pulse duration of 90 fs to 70 MW at pulse duration of 730 fs.

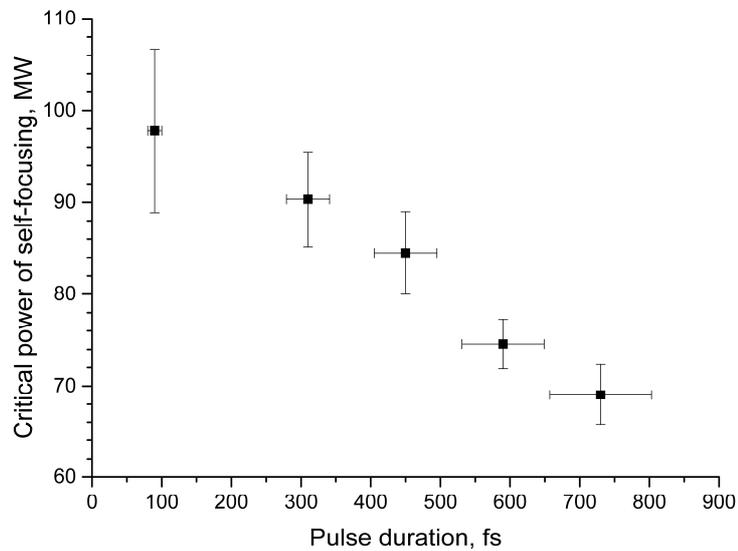

Fig.2. Dependence of the self-focusing critical power on UV pulse duration.

3. **Summary and conclusion**

Thus, we experimentally studied the dependence of self-focusing critical power on UV ultrashort pulse stretched in transparent optics with duration in the time-range from 90 to 730 fs. It was shown that this dependence is monotonically decreases almost linearly with the pulse duration rise. It was also observed that when the pulse power exceeds the critical one, the maximum of linear plasma distribution along the UV laser filament does not shifted in the direction opposite to the laser pulse propagation as in the case of IR laser filament, but remains at the geometrical focus.

**Acknowledgements**

This research was supported by the Russian Foundation for Basic Research (Project # 17-02-00722).